\documentclass[%
preprint,
 amsmath,amssymb,
 aps,
]{revtex4-1}

\usepackage{graphicx}
\usepackage{dcolumn}
\usepackage{bm}
\usepackage{amsmath,amssymb,graphicx}
\input epsf
\usepackage[utf8]{inputenc}
\usepackage[T1]{fontenc}
\usepackage{amsbsy}
\usepackage{latexsym}
\usepackage{color}

\usepackage{psfrag}

\usepackage{tikz}

\begin{document}
\title{Spacing Homogenization in Lamellar Eutectic Arrays with Anisotropic Interphase Boundaries}
\author{M. Ignacio, M. Plapp}
\affiliation{Laboratoire de Physique de la Matière Condensée, Ecole Polytechnique, CNRS, 91128 Palaiseau, France.}
\date{25 November 2019}

\begin{abstract}

We analyze the effect of interphase boundary anisotropy on the dynamics of lamellar
eutectic solidification fronts, in the limit that the lamellar spacing varies slowly
along the envelope of the front. In the isotropic case, it is known that the spacing
obeys a diffusion equation, which can be obtained theoretically by making two
assumptions: (i) the lamellae always grow normal to the large-scale envelope of
the front, and (ii) the Jackson-Hunt law that links lamellar spacing and front temperature
remains locally valid. For anisotropic boundaries, we replace hypothesis (i) by
the symmetric pattern approximation, which has recently been found to yield good
predictions for lamellar growth direction in presence of interphase anisotropy.
We obtain a generalized Jackson-Hunt law for tilted lamellae, and an evolution
equation for the envelope of the front. The latter contains a propagative term
if the initial lamellar array is tilted with respect to the direction of the
temperature gradient. However, the propagation velocity of the propagative wave modes 
are found to be small, so that the dynamics of the front can be reasonably described
by a diffusion equation with a diffusion coefficient that is modified with
respect to the isotropic case.

\end{abstract}

\maketitle

\section{Introduction}
\label{section:intro}
Eutectic alloys solidify into two-phase composite solids for a wide range of
compositions. The geometric structure of the composite is the result of a
pattern-formation process that takes place at the solid-liquid interface.
The patterns are shaped by the interplay between solute diffusion through the
liquid and capillary forces at the interfaces. This leads to the emergence
of lamellae if the volume fractions of the two phases are comparable. For
strongly different volume fractions, fibers of the minority phase inside
a matrix of the majority phase are found.

Eutectic solidification can be studied under well-controlled conditions by directional
solidification of thin samples \cite{Jackson_Hunt_TMSA_1966,Seetharaman_Trivedi_MTA_1988,
Ginibre_Faivre_PRE_1997}. In this geometry, most often the lamellar morphology 
emerges, and the crystallization front is quasi-one-dimensional.
In the absence of external perturbations and boundary effects, the lamellar 
pattern generally becomes more regular with time, that is, the lamellar spacing 
gets more and more homogeneous.

In a seminal paper, Jackson and Hunt have analyzed steady-state growth of
eutectic composites \cite{Jackson_Hunt_TMSA_1966}. They established
a relation between the average undercooling at the solidification
front $\Delta T$ -- the difference between the front temperature and
the eutectic temperature -- and the lamellar spacing $\lambda$. The curve
$\Delta T(\lambda)$ exhibits a single minimum; the spacings observed in
experiments on extended samples are typically distributed in a narrow range
around the spacing $\lambda_m$ that corresponds to this 
minimum \cite{Trivedi_Kurz_MTA_1991}.

Jackson and Hunt also qualitatively analyzed the stability of lamellar
arrays, under the hypothesis (which they attributed to Cahn) that lamellae
always grow in the direction that is perpendicular to the large-scale envelope of the
lamellar front. Then, in a (convex) bump of the front, the spacing gets
larger when solidification proceeds. If the undercooling increases with
the spacing, then the bump recedes in the temperature gradient and the front is stable; in contrast,
if the undercooling decreases with increasing spacing, the bump advances further
and the front is unstable. Eventually, the amplification of the front
deformation will lead to lamella pinchoff and elimination. The stability
of the front hence depends on the slope of the curve $\Delta T(\lambda)$.

These arguments were later formalized by Langer and
co-workers \cite{Langer_PRL_1980,Datye_Langer_PRB_1981}. They established
that the spacing obeys a diffusion equation, as is generally the case for
one-dimensional pattern-forming systems that exhibit a characteristic length
scale \cite{Manneville_1990,Cross_Hohenberg_RMP_1993}. The spacing diffusion
coefficient is proportional to $d\Delta T/d\lambda$ and becomes negative
for spacings smaller than $\lambda_m$. This means that the array is unstable for
spacings smaller than $\lambda_m$.

When experiments and numerical simulations became precise enough to directly
test these predictions, it was found that the normal growth hypothesis was
not exactly satisfied: the trijunctions also slightly move along the front
envelope, which gives an additional contribution to the spacing diffusion
coefficient that is always positive and hence stabilizing
\cite{Akamatsu_Karma_PRE_2002,Akamatsu_Karma_MMTA_2004}. 
Whereas it is likely that this lateral drift of the trijunctions originates from the
interaction of the diffusion field in the liquid, which depends on the local lamellar
spacing, and the shape of the solidification front at the scale of the individual
lamellae, no quantitative analytic expression for this contribution has been 
obtained so far. Instead, a single phenomenological parameter was fitted,
which could reproduce the results of both simulations and experiments.

All the theoretical analyses cited above neglect crystallographic effects, and assume
that all the interfaces are isotropic. This is also a standard assumption made in numerical
simulations \cite{Kassner_Misbah_PRA_1991,Karma_Sarkissian_MMTA_1996,Parisi_Plapp_AM_2008,Parisi_Plapp_EPL_2010}.
However, crystallographic effects are often important. This is obviously the case for
irregular eutectics, in which one or both of the solid-liquid interfaces are facetted. But
even in alloys where both solid-liquid interfaces are microscopically rough, crystallographic
effects can come into play through the solid-solid interfaces. In a eutectic grain, the
two solid phases have a fixed relative orientation with respect to each other, which can
differ between different grains. A distinction has been made between ``floating'' grains,
in which the solid-solid interfaces (interphase boundaries, IB) are isotropic, and ``locked''
grains, in which they are anisotropic and tend to follow certain crystallographic directions
\cite{Caroli_Mergy_JCG_1992}. In locked grains, lamellae can grow tilted with respect
to the direction of the temperature gradient, which clearly violates the hypothesis of
normal growth.

This behavior was recently studied in more detail by the new method of
{\em rotating directional solidification} \cite{Akamatsu_Faivre_AM_2012a}.
The results can be interpreted by taking into account the torque that is exerted on
the triple line by the anisotropy of the solid-solid interfaces. Instead of the interphase 
boundary itself, it is now the generalized surface tension vector 
$\vec\sigma$ \cite{Cahn_Hoffman_SS_1972}, which combines surface tension and torque, 
that is perpendicular to the front envelope. Since this entails that, in steady state, 
the solid-liquid interfaces have a mirror-symmetric shape with respect to the center
of each lamellae, this hypothesis was called {\em symmetric pattern approximation} 
(SPA) \cite{Akamatsu_Faivre_AM_2012b}. The SPA makes it possible to predict the
growth direction of the lamellae in steady state if the anisotropic surface free energy 
of the IB is known. Good agreement between the SPA and numerical simulations using 
boundary-integral and phase-field techniques was found \cite{Ghosh_Akamatsu_PRE_2015}. 

Here, we analyze how this torque alters the ``geometric part'' of the
spacing relaxation mechanism. In other words, we examine what is the
consequence of replacing Cahn's ansatz with the SPA. In a first step, we generalize
the Jackson-Hunt calculation, taking into account that in the presence of interphase 
boundary anisotropy the steady state is tilted. We demonstrate that the relation 
between undercooling and spacing keeps the same form, with the value of the 
minimum undercooling and the corresponding spacing depending on the tilting angle.

The tilt has a dramatic effect on the spacing dynamics because it
induces a breaking of the parity (right-left) symmetry in the base state.
In the case where the growth direction is aligned with an extremum of
the interphase boundary energy, the torque and thus the tilt angle are
zero. Then, the evolution equation for the lamellar spacing is again
a diffusion equation, but with a diffusion coefficient that is modified
by the interfacial anisotropy. In contrast, when the base state is
tilted, no closed-form evolution equation for the spacing can be written
down. Instead, an equation for the front shape can be formulated,
which is shown to have propagative solutions that can be damped or
amplified with time.

\section{Model}
\label{section:SM}

\subsection{Directional Solidification}
\label{section:LMS}
We consider the solidification of a binary eutectic alloy into two distinct solid phases called 
$\alpha$ and $\beta$. The sample is solidified by pulling it with a constant velocity $V$ from 
a hot to a cold zone; the externally imposed temperature gradient is aligned with the pulling
direction, and its magnitude is denoted by $G$. For a sufficiently thin sample, a two-dimensional
treatment is appropriate. The solid consists of a succession of pairs of lamellae of the phases $(\alpha,\beta)$.
In order to write down the system of equations ruling the evolution of the composition field, we assume that:
\begin{itemize}
 \item The molar densities of the solid and liquid phases are the same so that the total volume remains constant in time.
 \item Diffusion in the solid phases is neglected (one-sided model).
 \item Solute transport in the liquid is much slower than heat transport (\textit{i.e.} high Lewis number limit).
 \item Convection in the liquid is neglected (solute transport occurs only by diffusion). This is appropriate for a thin sample.
 \item Elasticity and plasticity in the solid phases are neglected.
 \item Heat conductivities are equal in all phases, so that the temperature field is independent of the shape of the solid-liquid interface.
 \item The latent heat rejected during solidification can be neglected.
  
\end{itemize}

Consequently (last two points), the temperature field is given by the {\em frozen temperature approximation},
\begin{equation}
 \label{FBP_1}
T(x,z,t) = T_E + G (z - V t),
\end{equation}
in the sample frame $\mathcal{R}_0(\hat{x}_0,\hat{z}_0)$, where $\hat z_0$ is the direction of the
pulling velocity and the temperature gradient. For convenience, we have chosen that the coordinate $z=0$ 
corresponds to the eutectic temperature $T_E$ at $t=0$.

\subsection{Free-Boundary Problem}
\label{section:GT}
Under the assumptions listed above, the fundamental free-boundary problem that describes
eutectic solidification is readily written down. In the liquid, the concentration field $C(x,z,t)$ obeys the
diffusion equation,
\begin{equation}
\frac{\partial C}{\partial t} = D \vec\nabla^2 C,
\end{equation}
with $D$ the solute diffusivity in the liquid. 
This equation has to be solved subject to the Gibbs-Thomson equation at the solid-liquid interface.
The shape of the solid-liquid interface is described by the function $z_{\rm int}(x,t)$; 
the interface undercooling is given by
\begin{equation}
 \label{GT_1}
 \Delta T = T_E - T(z_{\rm int}(x,t)) = \Delta T_D + \Delta T_c + \Delta T_k
\end{equation}
with $T_E$ the eutectic temperature, $ T(z_{\rm int}(x,t))$ the temperature at the solid/liquid interface,
and $\Delta T_D$, $\Delta T_c$, $\Delta T_k$ stand respectively for the diffusion, capillary 
and kinetic contributions.
The first term links the concentration at the interface to the interface temperature according to
\begin{equation}
  \label{eq_TD}
  \Delta T_D = - m_i (C_i(x,z_{\rm int}(x,t),t) -C_E), 
\end{equation}
with $m_i = \mathrm{d}T/\mathrm{d}C_i$ the liquidus slope of phase $i$, $C_i(x,_{\rm int})$ the concentration 
on the liquid side of the interface and $C_E$  the eutectic composition.
The term $\Delta T_c$ arises from the capillary force that shifts the melting point by an amount
that is proportional to the interface curvature $\kappa$ (we recall that we assume that the
solid-liquid interfaces are isotropic):
\begin{equation}
  \label{eq_TC}
  \Delta T_c = \frac{\gamma_{iL} T_E}{L_i} \kappa, 
\end{equation}
with $\gamma_{iL}$ the solid/liquid surface tension and $L_i$ the latent heat per unit volume for phase $i$.

Finally, the kinetic contribution reads
\begin{equation}
  \label{eq_Tk}
  \Delta T_k = \frac{V_n}{\mu_i}, 
\end{equation}
with $V_n$ the local velocity normal to the interface and  $\mu_i$ the linear kinetic coefficient (the
interface mobility).

The free-boundary problem is completed by the Young-Herring equation, to be discussed below,
and the Stefan condition that expresses the conservation of solute at the moving solid-liquid interface,
\begin{equation}
V_n\Delta C_{sl}^i = - \hat{n}\cdot D\vec\nabla C,
\end{equation}
where $\Delta C_{sl}^i$ is the concentration difference between the solid $i$ ($i=\alpha,\beta$) and the liquid,
$\hat{n}$ is the unit vector normal to the S/L interface, and $V_n$ the normal velocity of the interface. Since it turns 
out that the solid-liquid interfaces always remain close to the eutectic temperature for slow growth, it is a good 
approximation to set $\Delta C_{sl}^i$ equal to the equilibrium concentration differences at $T_E$.
 
Since the temperature field is set by Eq.~(\ref{FBP_1}), the interface position satisfies
\begin{equation}
 \label{eq_loceq}
  \Delta T = G (z_{\rm int}-V t).
\end{equation}
One can obtain a dimensionless formulation of the Gibbs Thomson law by defining
the dimensionless composition field
\begin{equation}
 \label{eq_u}
  c(x,z,t) = \frac{C(x,z,t)-C_E}{\Delta C},
\end{equation}
where $\Delta C= C^s_\beta -C^s_\alpha$ the eutectic plateau in the phase diagram, with $C_\alpha^s$ 
and $C_\beta^s$ the concentration of the solid phases.

Using Eqs.~(\ref{eq_loceq}) and (\ref{eq_u}), the dimensionless Gibbs Thomson law becomes (the minus sign is 
for the $\alpha-$phase, the plus for the $\beta-$phase),
\begin{equation}
 \label{GT_2}
 \frac{\Delta T}{|m_i| \Delta C}  = -\frac{z_{\rm int}-V t}{\ell_T^i} = \mp c_i(x,z_{\rm int}) + d_i \kappa + \beta_i V_n
\end{equation}
where
\begin{equation}
 \label{FBP_6}
 \ell_T^i = \frac{|m_i| \Delta C}{ G}
\end{equation}
are the thermal lengths,
\begin{equation}
 \label{FBP_7}
 d_i = \frac{\gamma_{iL} T_E }{|m_i| L_i \Delta C}
\end{equation}
are the capillary lengths, and
\begin{equation}
 \label{FBP_8}
 \beta_i = \frac{1}{|m_i| \mu_i \Delta C}
\end{equation}
are the kinetic coefficients.
We also introduce the diffusion length
\begin{equation}
 \label{FBP_9}
 \ell_{D} = \frac{D}{V}
\end{equation}
with $D$ the diffusion coefficient of the solute in the liquid phase.

For most metallic alloys and their organic analogs that have microscopically rough solid-liquid 
interfaces, the kinetic term $\Delta T_k$ can be neglected compared to both $\Delta T_D$ and 
$\Delta T_c$ \cite{Kramer_Tiller_JCP_1965}. This term will be dropped from now on.

\subsection{Young-Herring Equation}
\label{section:YHE}

As already mentioned above, the free-boundary problem is completed by the Young-Herring
equation, which is a statement of capillary force balance at the triple lines (triple points in
the quasi-two-dimensional approximation). Before stating it, let us make a few more comments
on the crystallography of eutectics.

For alloy systems with microscopically rough solid-liquid interfaces, the interface free energy
of the solid-liquid interfaces depends only weakly on the interface orientation -- it varies typically 
only by a few percent. Therefore, we will assume in this work that the solid-liquid interfaces are isotropic. 
In contrast, the solid-solid interfaces (interphase boundaries, IB) may be strongly anisotropic. A eutectic
composite consists of eutectic grains. In each grain, all the domains of a given phase ($\alpha$
or $\beta$) have the same orientation. The {\em relative} orientation of $\alpha$ and $\beta$
is therefore fixed in a given grain, but may vary between different grains. However, the IB can
still freely choose its orientation; therefore, an IB energy $\gamma_{\alpha\beta}(\hat{n}_{\alpha \beta})$ may 
be defined as a function of orientation. Here, $\hat{n}_{\alpha \beta}$ is the unit normal vector of the IB, which
is equivalent to two polar angles in three dimensions. However, in the quasi-two-dimensional approximation
for a thin sample, the IB are supposed to remain perpendicular to the sample walls, and therefore the IB can explore
only the orientations that lie within the sample plane. As a consequence, $\gamma_{\alpha\beta}$
is a function of a single angle $\phi$ (the polar angle in the sample plane), and the function
$\gamma_{\alpha\beta}(\phi)$ is the intersection of the full three-dimensional $\gamma$-plot and the
sample plane.

In the following, we denote by $\gamma_{\alpha\beta}(\phi)$ the orientation-dependent IB
energy {\em in the crystallographic frame}, that is, with respect to some reference axis of
the crystal. If the sample is rotated with respect to the laboratory frame by an angle $\phi_R$,
as can be done in the method of rotating directional solidification \cite{Akamatsu_Faivre_AM_2012a}, the
orientation-dependent IB energy {\em in the laboratory frame} will be given by $\gamma_{\alpha\beta}(\phi-\phi_R)$,
where $\phi$ is the angle between the IB orientation and the direction of the temperature
gradient. We choose that $\phi_R=0$ corresponds to a state in which a minimum of the
IB energy is aligned with the temperature gradient. As a generic example, we use $n$-fold
harmonic functions of the form
\begin{equation}
\gamma_{\alpha \beta}(\phi) = \gamma_0 [1 - \epsilon \cos(n(\phi - \phi_R))]
\label{eq:nfold}
\end{equation}
where $\epsilon$ is the anisotropy strength and $\gamma_0$ the average surfarce tension which be set to $1$ in the following. It should be mentioned that for a eutectic consisting 
of crystals with centrosymmetric unit cells, a two-fold symmetry ($n=2$) of the IB energy is always present.

\begin{figure*}[h!]
\begin{center}
\includegraphics[width=.8\textwidth]{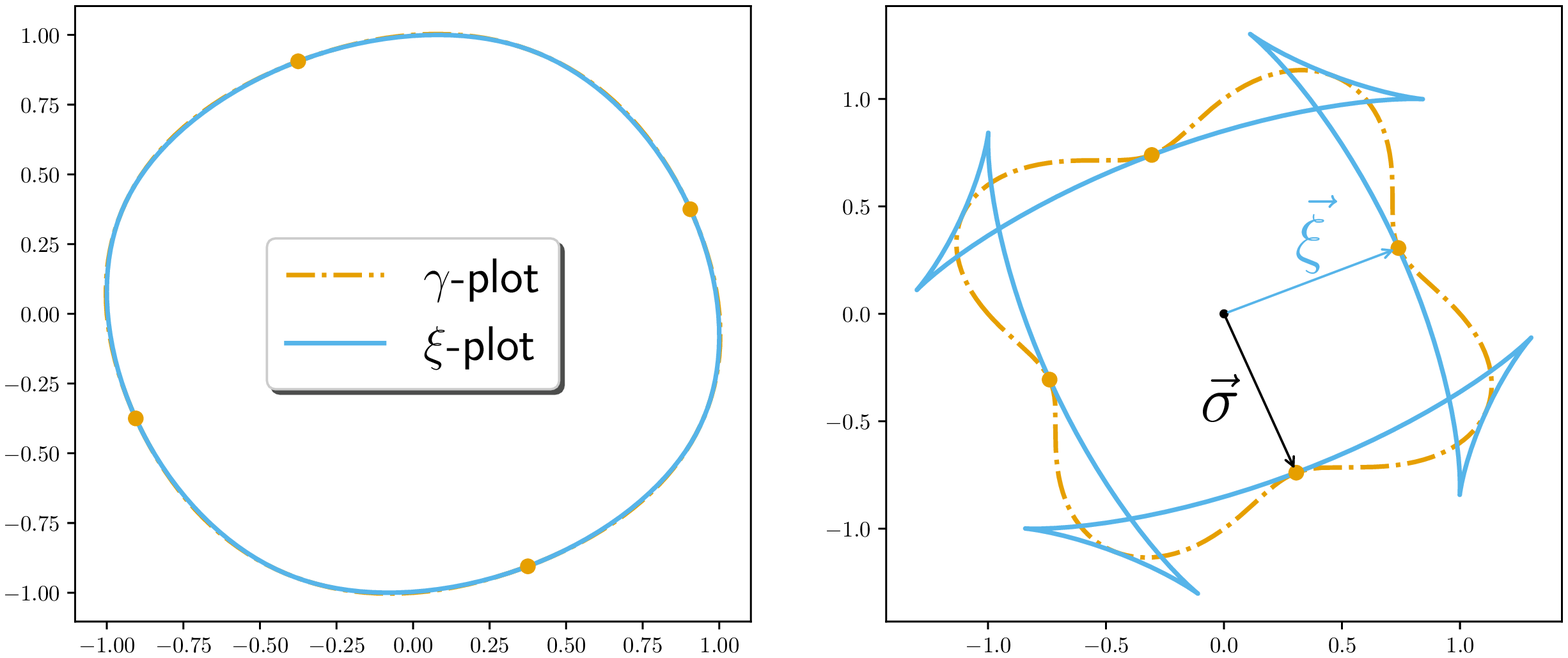}
\end{center}
\caption{$\gamma$-plot and $\xi$-plot of the interfacial free energy given by $\gamma_{\alpha \beta} = 1 - \epsilon \cos(n(\phi - \phi_R))$ with $\phi_R=\pi/8$ and $n=4$.
Left: $\epsilon=0.02$ (without missing orientation). Right: $\epsilon=0.2$ (with missing orientations).
The orange dots indicate the positions of the minima of $\gamma_{\alpha \beta}$.  }
\label{fig_1}
\end{figure*}

The force balance at trijunctions can be easily stated using the Cahn-Hoffman formalism \cite{Cahn_Hoffman_SS_1972,Wheeler_JSP_1999}.
Let $\hat{n}_{\alpha \beta}$ be the unit normal vector to the solid interphase and
$\hat{t}_{\alpha \beta} = -\mathrm{d} \hat{n}_{\alpha \beta}/\mathrm{d}\phi$ the unit tangential vector to the solid interphase.
With these definitions, the Cahn-Hoffman vector reads
\begin{equation}
 \label{eq_CHV}
  \vec{\xi}_{\alpha \beta} = \gamma_{\alpha \beta}(\phi) \hat{n}_{\alpha \beta} - \gamma_{\alpha \beta}'(\phi)\hat{t}_{\alpha \beta}.
\end{equation}
In addition, in two-dimension, one can define a unique generalized surface tension vector as
\begin{equation}
 \label{eq_STV}
  \vec{\sigma}_{\alpha \beta}	 =  \gamma_{\alpha \beta}(\phi) \hat{t}_{\alpha \beta} + \gamma_{\alpha \beta}'(\phi) \hat{n}_{\alpha \beta}.
\end{equation}
The equilibrium shape of a $\beta$ inclusion in an $\alpha$ crystal (or an $\alpha$ inclusion in a $\beta$
crystal) is given by the inner envelope of the polar plot of $\xi$ ($\xi-$plot). When the stiffness 
$\gamma_{\alpha \beta}+\gamma''_{\alpha \beta}\leq0$, a range of orientations is excluded from
the equilibrium shape (missing orientations, MO), which then exhibits sharp corners.
For a $n-$fold solid-solid interfacial free energy of the form given by Eq.~(\ref{eq:nfold}),
the condition for negative stiffness reads $\epsilon \geq (n^2-1)^{-1}$.
We display the polar plots of $\gamma$ ($\gamma-$plot) and $\xi$ ($\xi-$plot) in Fig.~\ref{fig_1}, 
for $\epsilon=0.05$ (without MO) and for $\epsilon=0.15$ (with MO) for a 4-fold interface free energy.

The vector $\vec{\sigma}_{\alpha \beta}$ gives the surface tension force. It allows us 
to write the Young-Herring equation at the trijunction
\begin{equation}
 \label{eq_YH}
   \vec{\gamma}_{\alpha \ell} + \vec{\gamma}_{\beta \ell}  + \vec{\sigma}_{\alpha \beta}(\phi) =0
\end{equation}
where $\vec{\gamma}_{\alpha \ell} =\gamma_{\alpha \ell} \hat{t}_{\alpha \ell}$ and $\vec{\gamma}_{\beta \ell} =\gamma_{\beta \ell} \hat{t}_{\beta \ell}$, with $\hat{t}_{\alpha \ell}$ and $\hat{t}_{\beta \ell}$, respectively the tangential unit vectors to the $\alpha-$ and $\beta-$liquid interfaces pointing away from the trijunction.

\section{Symmetric Pattern Approximation (SPA)}
\label{section:SPA}

\begin{figure*}[h!]
\begin{center}
\includegraphics[height=6cm]{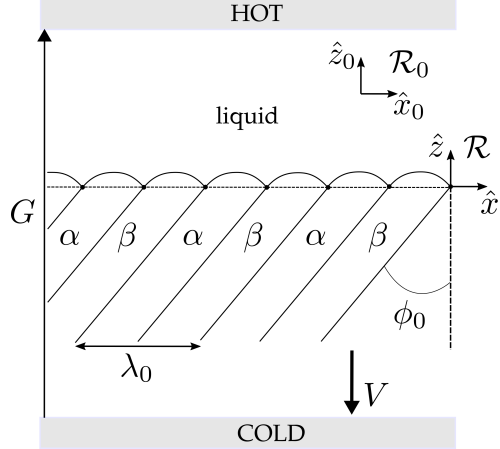}
\end{center}
\caption{Schematics of a tilted reference state $\{\lambda_0,\phi_0\}$ under directional solidification conditions.}
\label{fig_ab}
\end{figure*}

Consider a steady-state lamellar array with a regular spacing $\lambda_0$ (called ``undeformed 
state'' in the following). In presence of an interphase anisotropy, the solid interphase can exhibit 
a tilting angle $\phi_0$ with respect to the direction of the thermal gradient.
We introduce the frame of study $\mathcal{R}(\hat{x},\hat{z})$, moving at 
constant velocity $\vec{V}=V (\hat{z} + \tan \phi_0 \hat{x})$ with respect to the sample frame 
$\mathcal{R}_0(\hat{x}_0,\hat{z}_0)$.
This means that the
trijunction points drift laterally with a velocity $V_\parallel = V\tan\phi_0$, see Fig.~\ref{fig_ab} for the notations.

Experiments and numerical simulations show that the “heads” of the lamellae are approximately mirror symmetric
with respect to the midplane of the lamellae. Therefore, the contact angles (the angles between the solid-liquid
interfaces direction of the isotherms $\hat x$) of the solid-liquid interfaces at the trijunctions are also 
approximately the same on both sides of a lamella. This is only possible if the surface tension vector 
$\vec{\sigma}_{\alpha \beta}$ is approximately perpendicular to the envelope of the solid-liquid front.
The assumption that $\vec\sigma_{\alpha\beta}$ is {\em exactly} perpendicular to the front
was called in Ref.~\cite{Akamatsu_Faivre_AM_2012b} the {\em symmetric pattern approximation} (SPA).
Introducing the unit vectors parallel $\hat{t}_f$ and perpendicular $\hat{n}_f$ to the large-scale solid-liquid front, 
the SPA reads
\begin{equation}
 \label{eq_SPA}
 \vec{\sigma}_{\alpha \beta} \cdot \hat{t}_f = 0.
\end{equation}

\begin{figure*}[h!]
\begin{center}
\includegraphics[width=.4\textwidth]{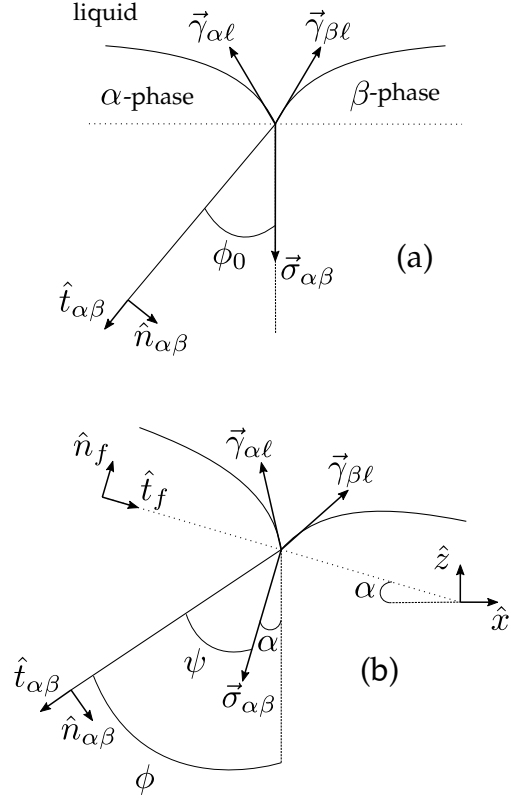}
\end{center}
\caption{Schematics of the angles and vectors using the Symmetric Pattern Approximation (SPA) for a) a planar and b) a tilted S/L fronts. }
\label{fig_angle0}
\end{figure*}

Consequently, the Young-Herring condition Eq.~(\ref{eq_YH}) expressed in the basis formed by $\{\hat{t}_f,\hat{n}_f\}$ reads
\begin{eqnarray}
 \label{eq_YH0}
    \gamma_{\alpha \ell} \cos(\theta_\alpha) - \gamma_{\beta \ell} \cos(\theta_\beta) &=& 0,  \\ \nonumber
    \gamma_{\alpha \ell} \sin(\theta_\alpha) + \gamma_{\beta \ell} \sin(\theta_\beta) &=&  |\vec{\sigma}_{\alpha \beta}(\phi_0-\phi_R)|. 
\end{eqnarray}
where $\theta_\alpha$ and $\theta_\beta$ are the contact angles, both taken positive.
In the following, we investigate the consequences of the SPA first for a front that is perpendicular to the
pulling direction, and then for a tilted solid-liquid front, see Fig.~\ref{fig_angle0}.

\subsection{Base State: Front Perpendicular to the Growth Direction}
\label{SPA_P}

For an undeformed steady-state, see Fig.~\ref{fig_angle0}(a), the SPA given by Eq.~(\ref{eq_SPA}) leads to 
\begin{equation}
 \label{eq_tilt0}
  \phi_0 = - \arctan\left(\frac{\gamma_{\alpha \beta}'(\phi_0 -\phi_R)}{\gamma_{\alpha \beta}(\phi_0-\phi_R)}\right).
\end{equation}
The solution of Eq.~(\ref{eq_tilt0}) gives the steady-state tilt angle as a function of the orientation of the bicrystal, $\phi_R$.
It is worth noting that Eq.~(\ref{eq_tilt0}) has one single solution if the stiffness $\gamma_{\alpha \beta}+\gamma''_{\alpha \beta}>0$ for all orientations, and can have up to three solutions if $\gamma_{\alpha \beta}+\gamma''_{\alpha \beta}<0$
for some range of orientations. One may distinguish stable, meta-stable and unstable branches that can be associated to the
features of the $\xi$-plot \cite{Cabrera_SS_1964, Philippe_Plapp_JCG_2018}, see appendix \ref{APP_B} for details. For cases with MO, the system will select one of the two stable branches for a fixed $\phi_R$.
In contrast, if the system is brought to an initial state located on the unstable branch, we expect that the Herring 
instability \cite{Herring_PR_1951} will appear.

\begin{figure*}[h!]
\begin{center}
\includegraphics[width=.7\textwidth]{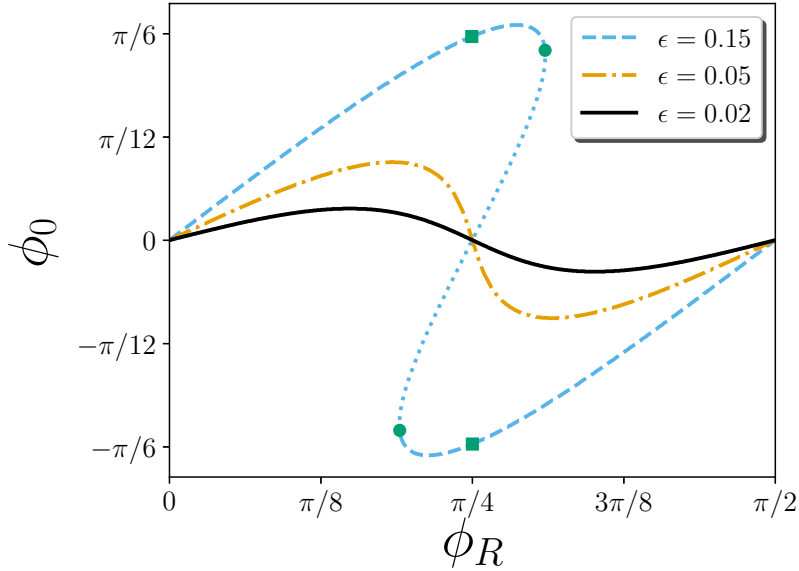}
\end{center}
\caption{Steady-state tilting angle $\phi_0$ as a function of the rotation angle $\phi_R$ within the SPA given by Eq.~(\ref{eq_tilt0}).
The solid/solid surface tension corresponds to $\gamma_{\alpha \beta} = 1 - \epsilon \cos(4(\phi_0 - \phi_R))$ with $\epsilon=0.02$, $0.05$ and $0.15$. For $\epsilon=0.15$, the dashed curves correspond to the stable and the metastable branches. The green $\blacksquare$ indicate the beginning of the metastable branches. The dotted curve corresponds to the unstable branch. The limit between metastable and unstable branches is marked by the green $\bullet$.}
\label{fig_phi0R}
\end{figure*}

In figure \ref{fig_phi0R}, we plot the solution of Eq.~(\ref{eq_tilt0}) for $\phi_0$ as a function of $\phi_R$ for a surface energy with $4$-fold symmetry with $\epsilon=0.02, 0.05$ (without MO) and $0.15$ (with MO). The symbols indicate the limit of the metastable branches. The amplitude of the variation of $\phi_0$ always increases with $\epsilon$. 

Furthermore, applying the transformation $\phi_R \rightarrow \phi_R +\delta \phi_R$ and $\phi_0 \rightarrow \phi_0 +\delta \phi_0$,  Eq.~(\ref{eq_tilt0}) leads to the relation
\begin{equation}
 \label{eq_undef1}
 \delta \phi_0 = \frac{\gamma''_{\alpha \beta} \gamma_{\alpha \beta} - \gamma^{'2}_{\alpha \beta}}{\gamma_{\alpha \beta}(\gamma_{\alpha \beta} +\gamma''_{\alpha \beta})} \delta \phi_R. 
\end{equation}

\begin{figure*}[h!]
\begin{center}
\includegraphics[width=.7\textwidth]{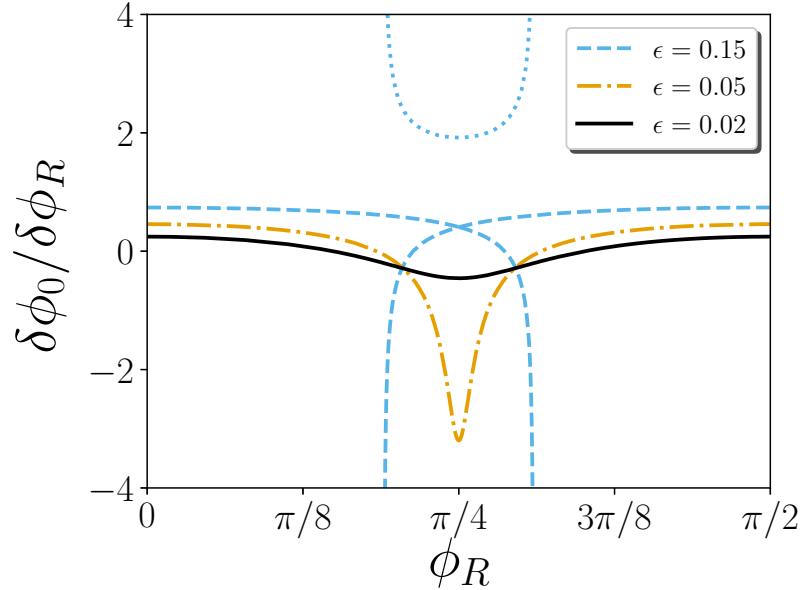}
\end{center}
\caption{Evolution of the slope $\delta \phi_0 / \delta \phi_R$ with respect to the crystallographic angle $\phi_R$ within the SPA given by Eq.~(\ref{eq_undef1}).
The S/S surface tension corresponds to $\gamma_{\alpha \beta} = 1 - \epsilon \cos(4(\phi_0 - \phi_R))$ with $\epsilon=0.02$, $0.05$ and $0.15$. The dotted curves give the metastable and the unstable branches for $\epsilon=0.15$.}
\label{fig_derphi0R}
\end{figure*}

For cases without MO,
according to Eq.~(\ref{eq_undef1}), the values of $\phi_R$ corresponding to a sign change of the
slope of the curve $\phi_0(\phi_R)$ are given by the solutions of $\gamma''_{\alpha \beta} \gamma_{\alpha \beta} - \gamma^{'2}_{\alpha \beta}=0$, see Fig.~\ref{fig_derphi0R}.
Close to a mininum of anisotropy (\textit{i.e.} $\phi_R$ mod$[2\pi/n] =0$), the sign of $\delta \phi_0/\delta \phi_R$ is positive which means that the tilting angle $\phi_0$ tends to ``follow'' the rotation $\delta \phi_R$.
Conversely, around a maximum of anisotropy (\textit{i.e.} $\phi_R$ mod$[2\pi/n] =\pi/n$), the slope of $\phi_0(\phi_R)$ is negative, and therefore, any change of of the crystallographic angle $\delta \phi_R$ will lead to a change of
the tilting angle in the opposite direction.

\subsection{Inclined Front}
\label{SPA_I}

Considering an inclined planar front, see Fig.~\ref{fig_angle0}(b), characterized by the angle $\alpha$ between $\hat z$ and $\hat{n}_f$, the unit vectors normal and tangential to the front read
\begin{eqnarray}
 \label{eq_unitntf}
  \hat{n}_{f} &=& -\sin(\alpha) \hat{x} + \cos(\alpha) \hat{z}, \\ \nonumber
  \hat{t}_{f} &=& \cos(\alpha) \hat{x} + \sin(\alpha) \hat{z}.
\end{eqnarray}

The coordinates of the interphase unit vectors in the frame $\mathcal{R}(\hat{x},\hat{z})$ read
\begin{eqnarray}
 \label{eq_unitnt}
  \hat{n}_{\alpha \beta} &=& \cos(\phi) \hat{x} + \sin(\phi) \hat{z}, \\ \nonumber
  \hat{t}_{\alpha \beta} &=&  -\frac{\mathrm{d} \hat{n}_{\alpha \beta}}{\mathrm{d}\phi} = \sin(\phi) \hat{x} - \cos(\phi) \hat{z}.
\end{eqnarray}

The SPA reads
\begin{equation}
 \label{eq_SPA2}
 \hat{\sigma}_{\alpha \beta} \cdot \hat{t}_f = 0 \Rightarrow -\tan \alpha = \frac{\gamma_{\alpha \beta} \tan \phi + \gamma'_{\alpha \beta}}{\gamma'_{\alpha \beta} \tan \phi - \gamma_{\alpha \beta}}
\end{equation}
Introducing the angle $\psi$ between the solid interphase and $\vec{\sigma}_{\alpha \beta}$, such as $\tan \psi = - \gamma'_{\alpha \beta}/\gamma_{\alpha \beta}$, inside Eq.~(\ref{eq_SPA}) leads to the simple 
geometric relation
\begin{equation}
 \label{eq_tilt}
  \alpha = \phi - \psi. 
\end{equation}
In addition, for the isotropic case (\textit{i.e.} $\gamma_{\alpha \beta}'=0$), one gets $\alpha = \phi$ which corresponds to Cahn's ansatz
(the solid interphase remains perpendicular to the solid/liquid front during growth). 

Applying the transformation $\phi \rightarrow \phi_0 +\delta \phi$ and
$\alpha \rightarrow \alpha_0 +\delta \alpha$ (with $\alpha_0=0$), inside
Eq.~(\ref{eq_SPA2}), one obtains

\begin{eqnarray}
 \label{eq_def0}
   \delta \phi &=& \left( 1 - \left.\frac{\mathrm{d} \psi  }{ \mathrm{d} \phi }\right|_{\phi_0}  \right)^{-1}  \delta \alpha , \\ \nonumber
  			   &=& A_{SPA}(\phi_R ) \delta \alpha,
\end{eqnarray}
where the anisotropy function $ A_{SPA}(\phi_R )$ within the SPA is given by
\begin{equation}
 \label{eq_Aphi}
  A_{SPA}(\phi_R )= \left[ 1 + \left(\frac{\gamma'_{\alpha \beta}}{\gamma_{\alpha \beta}}\right)^2  \right] \frac{\gamma_{\alpha \beta}}{\gamma_{\alpha \beta}+\gamma''_{\alpha \beta}}.
\end{equation}
The anisotropy function corresponds to the proportionality factor between the variation
of the tilting angle $\phi$ and the variation of the angle $\alpha$ characterizing locally the deformation of the S/L front.

Some interesting points should be noted:
\begin{itemize}
 \item The sign of the anisotropy function $A_{SPA}(\phi_R )$ is imposed by the sign of the stiffness $\gamma_{\alpha \beta}+\gamma''_{\alpha \beta}$.
 \item The equation $A_{SPA}(\phi_R )=1$ has solutions for $\frac{\mathrm{d}\psi}{\mathrm{d}\phi}=0$ or equivalently
 $  \gamma''_{\alpha \beta} - (\gamma'_{\alpha \beta}/\gamma_{\alpha \beta})^2=0 $. 
 It has one trivial solution corresponding to the isotropic case ($\gamma_{\alpha \beta}$ constant).
 \item The function $A_{SPA}(\phi_R )$ diverges when $\frac{\mathrm{d}\psi}{\mathrm{d}\phi}=1$, 
 which corresponds to the case where the stiffness $\gamma_{\alpha \beta}+\gamma''_{\alpha \beta}$ equals $0$.
 \item For positive stiffness, the maxima of $A_{SPA}(\phi_R )$ are solutions of the equation $\frac{\mathrm{d}^2\psi}{\mathrm{d}\phi^2}=0$. In addition, using Eq.~(\ref{eq:nfold}),  the minimum and maximum values of the anisotropic function are $A_{SPA}^{min} = (1-\frac{n^2\epsilon}{1-\epsilon})^{-1}$ and
$A_{SPA}^{max} = (1+\frac{n^2\epsilon}{1+\epsilon})^{-1}$.
\end{itemize}

\begin{figure*}[h!]
\begin{center}
\includegraphics[width=.7\textwidth]{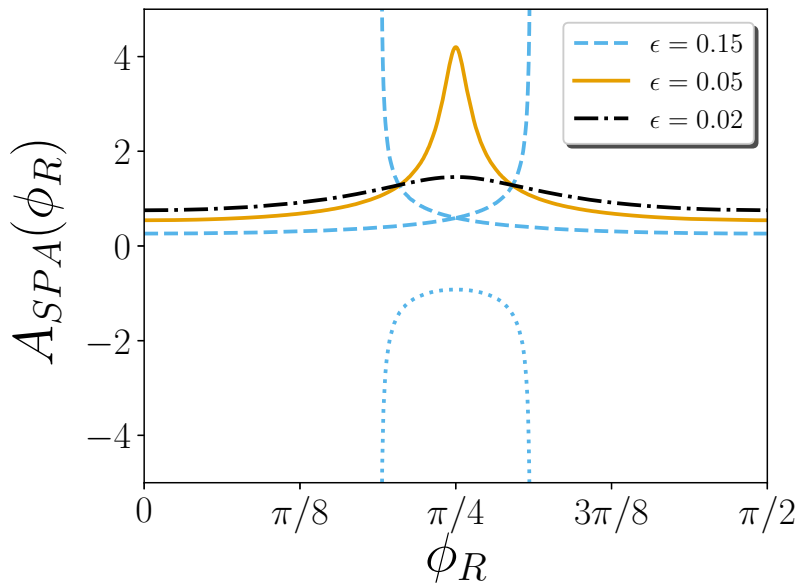}
\end{center}
\caption{Evolution of the anisotropy function $A_{SPA}(\phi_R )$ using $\gamma_{\alpha \beta} = 1 - \epsilon \cos(4(\phi_0 - \phi_R))$ with
 $\epsilon=0.02$, $0.05$ and $0.15$. The dotted curve gives the unstable branch for $\epsilon=0.15$. 
}
\label{fig_A}
\end{figure*}

We illustrate the behavior of the anisotropy function $A_{SPA}(\phi_R )$ for an interface energy with 4-fold 
symmetry in Fig.~\ref{fig_A}.

\section{Jackson Hunt Law with Interphase Anisotropy}
\label{section:JHLIA}

The analytical solution of the free-boundary problem with the real interface
shape is not known. In the Jackson-Hunt theory, the diffusion field is
calculated using a simplified interface shape, namely, a planar front. Moreover, 
the contributions to the interface undercooling are averaged over individual
lamella. 

In the moving frame $\mathcal{R}$, the average position of the
interface is given by
$\zeta_0 = \frac{1}{\lambda_0}\int_0^{\lambda_0} (z_{\rm int}(x)-V t) \mathrm{d}x$.
We introduce average quantities over one pair of lamellae, such as
\begin{equation}
 \label{eq_Average}
  \langle \dots \rangle =
  \left\{ \begin{array}{l}
  \frac{1}{\eta \lambda_0}\int_0^{\eta \lambda_0} \dots \mathrm{d}x  ~~\mbox{ for $\alpha-$phase} \\
  \frac{1}{(1-\eta) \lambda_0}\int_{\eta \lambda_0}^{\lambda_0} \dots \mathrm{d}x  ~~\mbox{ for $\beta-$phase} \\
\end{array}\right.
\end{equation}
with $\eta$ the nominal volume fraction of the $\alpha$ phase at the
eutectic temperature, which is related to $c_\infty$, the reduced composition
of the melt infinitely far ahead of the solidification front, by
$c_\infty=\eta c_\alpha +(1-\eta) c_\beta$.

The expression of the capillary contribution to the undercooling, given by the average of
Eq.~(\ref{eq_TC}), $\langle\Delta T_{c,i}\rangle \propto \langle  \kappa_i \rangle$, is directly obtained by 
averaging the local curvature $\kappa= -\partial_{xx}z_{\rm int}/(1+\partial_x{z_{\rm int}}^2)^{3/2}$. 
One gets
\begin{equation}
 \label{eq_avcurv}
 \langle\kappa\rangle =
\left\{ \begin{array}{l}
 \frac{2}{\eta \lambda_0} \sin(\theta_\alpha(\phi_0)) ~~\mbox{for $\alpha$-phase}\\ 
 \frac{2}{(1-\eta) \lambda_0} \sin(\theta_\beta(\phi_0)) ~~\mbox{for $\beta$-phase}\\ 
\end{array}\right.
\end{equation}
where $\theta_i$, the contact angles at the trijunction points, are fixed by the 
equilibrium condition of the capillary forces at the trijunctions (Young Herring equation), 
Eq.~(\ref{eq_YH0}). Note that in the SPA the contact angles on the two sides of each
lamellae are identical, even for a tilted steady state.

In order to calculate the average of the diffusion term, given by Eq.~(\ref{eq_TD}), $\langle \Delta T_{D,i} \rangle$, 
one assumes a flat S/L interface \cite{Datye_Langer_PRB_1981,Langer_PRL_1980}
(\textit{i.e.}, herein, the curvature can be seen as a perturbation).
In steady
state and in the frame of reference $\mathcal{R}$, the diffusion equation for the concentration field reads \cite{Kassner_Misbah_PRA_1992},
\begin{equation}
 \label{eq_diffeq}
  \nabla^2 c +\frac{1}{\ell_{D}} \left(\frac{\partial c}{\partial z}+\tan \phi_0\frac{\partial c}{\partial x}\right) = 0.
\end{equation}

In addition, for a planar S/L front perpendicular to the temperature gradient, the condition of mass conservation at the interface
(Stefan's condition \cite{Stefan_1889}) imposes
\begin{equation}
\label{eq_contiF}
\left. \frac{\partial c}{\partial z}\right|_{\zeta_0} = - \frac{1}{\ell_D} [c(x,\zeta_0) - c^s_i],
\end{equation}
with $c_i^s$ the reduced concentration of the solid phase $i$ ($i=\alpha,\beta$).

Let us now proceed to the solution of the diffusion equation.
For simplicity of notations, we will set $\zeta_0=0$ in the following
(that is, $z=0$ corresponds to the interface position). The general solution
of Eq.~(\ref{eq_diffeq}) for a system with the spatial periodicity $\lambda_0$ on $x$ reads
\begin{equation}
 \label{eq_geS}
  c(x,z) = c_{\infty} + \sum_{n=-\infty}^{+\infty} B_n \exp(-Q_n z) \exp( i k_n x)
\end{equation}
with $k_n = 2\pi n/\lambda_0$ the wave number of the mode $n$.
Inserting the general solution inside Eq.~(\ref{eq_diffeq}) and
keeping only the positive root for $Q_n$ (since the concentration field must tend towards
a constant far from the front), one gets
\begin{equation}
 \label{eq_Qn0}
  Q_n = \frac{1}{2\ell_D}\left(1 + \sqrt{ 1 + 4 \ell_D^2 k_n^2\left[1 - i \frac{ \tan \phi_0}{k_n \ell_D}\right] }\right).
\end{equation}
The Peclet number is introduced as the ratio of the lamellar
spacing and the diffusion length, $Pe=\lambda_0/\ell_D$. In the limit of small Peclet number
(that is, for slow growth),
$Pe \sim (|k_n|\ell_{D})^{-1} \ll1$ for $n\neq0$,
Eq.~(\ref{eq_Qn0}) can be simplified by keeping only the terms up to the first order in $Pe$, which yields
\begin{eqnarray}
\label{eq_Qn}
Q_n &\approx& \frac{1}{2\ell_{D}} + |k_n|\exp\left(-\frac{i \tan \phi_0}{2\ell_{D} k_n}\right) \nonumber \\
      &\approx& \frac{1}{2\ell_{D}} + |k_n| -i  \frac{\tan \phi_0}{2\ell_{D}} sign(n) .
\end{eqnarray}
Then, inserting Eq.~(\ref{eq_geS}) inside the continuity equation Eq.~(\ref{eq_contiF}) and integrating over $x$ (from $x=0$
to $x=\eta\lambda_0$ for $\alpha$ and from $x=\eta \lambda_0$ to $x={\lambda_0}$ for $\beta$), one obtains
the coefficients $B_n$ \cite{Datye_Langer_PRB_1981},
\begin{eqnarray}
 \label{eq_Bn}
  B_n &=& \frac{2\exp(i \eta k_n\lambda_0/2) \sin(\eta k_n \lambda_0 /2)}{\ell_D\lambda_0 k_n(Q_n -1/\ell_D)}, ~~ (\forall n)\\ \nonumber
      & \approx &  \frac{2\exp(i \eta k_n\lambda_0/2) \sin(\eta k_n \lambda_0 /2)}{\ell_D\lambda_0 k_n |k_n|}, ~~ (n \neq 0).
\end{eqnarray}
at the $0^{th}$ order in $Pe$.
The only difference with respect to the original Jackson-Hunt calculation is the presence of the imaginary part in the expression of $Q_n$ in Eq.~(\ref{eq_Qn}) which 
produces oscillations in the $z$ direction on the typical length $2\ell_D/\tan \phi_0$.
Those oscillations can be understood by realizing that the concentration
field is created by the rejection and absorption of solute at the moving interface. Since the distribution
of the sources and sinks drifts laterally along the front, the flux lines are slightly inclined with respect
to the solution for non-tilted growth. We have checked that the inclination angle is vanishingly small
in the small Peclet number regime.

The calculation of the average composition in front of each lamella yields
\begin{equation}
 \label{eq_AVa}
  \langle c_\alpha \rangle = \frac{1}{\eta \lambda_0}\int_0^{\eta \lambda_0}c(x,\zeta_0) \mathrm{d}x = c_\infty + B_0 + \frac{\lambda_0}{\ell_{D} \eta} P(\eta),
\end{equation}
\begin{equation}
 \label{eq_AVb}
  \langle c_\beta \rangle = \frac{1}{(1-\eta) \lambda_0}\int_{\eta \lambda_0}^{\lambda_0}c(x,\zeta_0) \mathrm{d}x = c_\infty + B_0 - \frac{\lambda_0}{\ell_{D} (1-\eta)} P(\eta),
\end{equation}
with
\begin{equation}
 \label{eq_Peta}
  P(\eta) = \sum_{n=1}^{\infty}\frac{\sin^2(\pi \eta n)}{(n\pi)^3}.
\end{equation}
This is the same result as for nontilted growth. This fact is not surprising since, for a planar interface,
no coupling between the interface shape and the lateral diffusion fluxes can occur \cite{Akamatsu_Faivre_AM_2012a}.
From the average compositions, one directly deduces the average diffusion undercooling 
$\langle \Delta T_{D,i} \rangle = -  \Delta C  m_i \langle c_i\rangle$.
It is worth noting that at this stage, the Fourier coefficient $B_0$, which corresponds 
to the amplitude of a uniform boundary layer of thickness $\ell_D$ moving ahead of the front, 
remains undetermined. The problem is closed by assuming that neighboring lamellae are at the same
temperature $\langle \Delta T_{\alpha} \rangle = \langle \Delta T_{\beta} \rangle$, which allows 
to determine the average undercooling  without knowing the analytical form of $B_0$.

From this, we deduce the Jackson-Hunt law in presence of anisotropic interphases
\begin{equation}
 \label{Gzetaa}
 \Delta T(\lambda_0,\phi_0)  = V  K_1 \lambda_0 + K_2(\phi_0) \lambda_0^{-1}.
\end{equation}
with
\begin{equation}
 \label{eq_K1}
  K_1 = \frac{\bar{m} P(\eta)}{D}  \frac{\Delta C}{\eta (1-\eta)}
\end{equation}
and
\begin{equation}
\label{eq_K2}
 K_2(\phi_0) = 2 \bar{m} \Delta C \left( \frac{d_\alpha \sin(\theta_\alpha(\phi_0))}{\eta} + \frac{d_\beta \sin(\theta_\beta(\phi_0))}{(1-\eta)}   \right)
\end{equation}
with
\begin{equation}
\label{eq_mbar}
\frac{1}{\bar{m}} = \frac{1}{|m_\alpha|} + \frac{1}{|m_\beta|}.
\end{equation}

Equivalently, introducing $\lambda_m(\phi_0) = \sqrt{K_2(\phi_0)/(V K_1)}$ and the minimum undercooling 
$\Delta T_m(\phi_0) = 2 \sqrt{V K_1 K_2(\phi_0)}=2V K_1 \lambda_m(\phi_0)$,
one has
\begin{equation}
 \label{Gzetaa2}
 \Delta T(\lambda_0,\phi_0) = \frac{\Delta T_m(\phi_0)}{2} \left( \frac{\lambda_0}{\lambda_m(\phi_0)} + \frac{\lambda_m(\phi_0)}{\lambda_0} \right).
\end{equation}

It is instructive to consider as an example a fictitious eutectic alloy with symmetric phase
diagram and identical surface tensions for the two solid-liquid interfaces, at the eutectic composition.
Indeed, the above expressions can be further simplified in that case: we have $\eta=1/2$, 
$|m_\alpha|=|m_\beta|=m$, $\theta_\alpha=\theta_\beta=\theta_\ell$;
the expression for $K_1$ reduces to $K_1 = 2 P(1/2) m \Delta C/D$, with $P(\eta=1/2) \approx 0.0339$.

The Young Herring law yields 
\begin{equation}
 \label{eq_YH1}
 \sin(\theta_\ell) = \frac{1}{2 \gamma_\ell} |\vec{\sigma}_{\alpha \beta}(\phi_0-\phi_R)|.
\end{equation}
The function $K_2$ given by Eq.~(\ref{eq_K2}) becomes $K_2(\phi_0) = 4 m \Delta C d_\ell \sin(\theta_\ell(\phi_0))$
with $ d_\alpha =d_\beta=d_\ell $ (the capillary length).
One directly obtains the expressions
\begin{equation}
 \label{eq_LJHsym}
 \lambda_m(\phi_0) = \sqrt{ \frac{2 d_\ell \ell_D \sin(\theta_\ell(\phi_0))}{P(1/2)}}
\end{equation}
and
\begin{equation}
 \label{eq_DTJHsym}
 \Delta T_m(\phi_0) = 4\sqrt{2} m\Delta C \sqrt{ \frac{d_\ell \sin(\theta_\ell(\phi_0))P(1/2)}{\ell_D}}.
\end{equation}
It turns out that $\Delta T_m$ and $\lambda_m$ are proportional to $|\vec{\sigma}_{\alpha \beta}(\phi_0-\phi_R)|^{1/2}$ according to
\begin{eqnarray}
 \label{eq_eqDTJH}
   \frac{\Delta T_m(\phi_0)}{m\Delta C} &=& 4 P(1/2) \frac{\lambda_m(\phi_0)}{\ell_D} , \\ \nonumber
  			   &=& 4 \sqrt{\frac{d_\ell P(1/2)}{\gamma_\ell \ell_D}}|\vec{\sigma}_{\alpha \beta}(\phi_0-\phi_R)|^{1/2}.
\end{eqnarray}

This fact can be easily understood: when $|\vec{\sigma}_{\alpha \beta}(\phi_0-\phi_R)|$ increases, the wetting angle $\theta_\ell$, and obviously the average curvature $\langle  \kappa \rangle$, has to increase as well to maintain the equilibrium of the capillary force at the trijunctions.
Therefore, the capillary contribution in the Gibbs Thomson law becomes stronger, which shifts $\Delta T_m$ and $\lambda_m$ towards higher values.

\begin{figure*}[h!]
\begin{center}
\includegraphics[width=.7\textwidth]{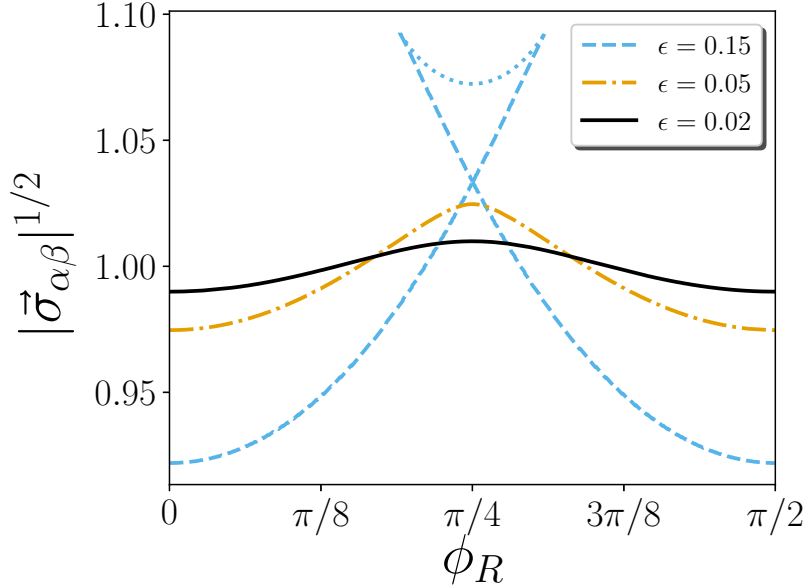}
\end{center}
\caption{Curve of $|\vec{\sigma}_{\alpha \beta}(\phi_0-\phi_R)|^{1/2}$ versus $\phi_R$ for a symmetric phase diagram.
We recall that both $\Delta T_m(\phi_0)$ and $\lambda_m(\phi_0)$ are proportional to $|\vec{\sigma}_{\alpha \beta}|^{1/2}$.
The IB energy is $\gamma_{\alpha \beta} = 1 - \epsilon \cos(4(\phi_0 - \phi_R))$ for anisotropy strengths $\epsilon=0.02$, $0.05$ and $0.15$. For $\epsilon=0.15$, the dashed curves correspond to the stable and the metastable branches, and the dotted curve to the unstable branch. }
\label{JH_sq}
\end{figure*}

We plot in Fig.~\ref{JH_sq} the function $|\vec{\sigma}_{\alpha \beta}(\phi_0-\phi_R)|^{1/2}$ versus $\phi_R$ using 
an IB energy of $\gamma_{\alpha \beta} = 1 - \epsilon \cos(4(\phi_0 - \phi_R))$. 

\section{Evolution of Modulated Fronts}
\label{section:EFSE}

We now wish to examine the time evolution of large-scale fronts for which the spacing and the
inclination of the lamellae may vary with position and time. We restrict our attention to weakly
modulated fronts, which can be described as perturbed steady-state fronts. As in previous
works, the deformation of the front with respect to the steady state may be described by
the displacements of the trijunction points, both parallel ($ \delta_x$) and perpendicular 
($\delta_z$) to the steady-state front. The large-scale envelope of the front (that is, a smooth
curve that is an interpolation of the average position $\zeta_0$ of each individual lamella)
is denoted by $\zeta(x,t)$ in the moving frame  $\mathcal{R}(\hat{x},\hat{z})$. When the displacements of the trijunction points 
vary slowly along the front, it is possible to take a continuum approach, in which the lateral 
displacements, the lamellar spacings and the tilting angles can be represented by slowly varying functions
of $x$ and $t$ denoted here by $\delta_x(x,t)$, $\lambda(x,t)$ and $\phi(x,t)$, respectively.
The local orientation of the front envelope is described by the unit normal vector 
$\hat{n}_f \propto (- \partial_x \zeta, 1)$. We introduce the angle $\alpha = -\arctan(\partial_x \zeta) \approx \partial_x \zeta$ between $z$ and $\hat{n}_f$.

\subsection{Fundamental Equations}
\label{section:FE}

In the long-wavelength limit, $ |\partial_x\lambda| / \lambda_0 \ll 1$, the local spacing $\lambda(x,t)$ 
of the deformed state is written as $\lambda(x,t) = \lambda_0 + \delta \lambda(x,t)$, 
which can be rewritten \cite{Datye_Langer_PRB_1981,Langer_PRL_1980}
\begin{equation}
 \label{lambdada}
  \lambda(x,t) \approx \lambda_0 \left(1 + \frac{\partial \delta_x}{ \partial x}\right).
\end{equation}
This of course implies
\begin{equation}
\frac{\partial \lambda}{\partial t}=\lambda_0 \frac{\partial}{\partial x}\frac{\partial \delta_x}{\partial t}.
\label{eq:dldt}
\end{equation}
In the following, we will assume that all the functions verify the theorem of Schwarz, such that the order 
of the derivatives with respect to $x$ and $t$ can be inverted.
Furthermore, we assume that the generalized Jackson Hunt law, Eq.~(\ref{Gzetaa2}), remains locally valid for a
smoothly varying spacing. Then, the undercooling at the S/L interface reads
\begin{equation}
 \label{eq:zeta}
-\zeta(x,t)  G= \frac{\Delta T_m(\phi(x,t))}{2} \left( \frac{\lambda(x,t)}{\lambda_m(\phi)} + \frac{\lambda_m(\phi)}{\lambda(x,t)} \right).
\end{equation}

The evolutions of $\delta_x$ and $\zeta$ are linked by the SPA. Indeed, if the local inclination of
the lamellae changes, this modifies the lateral drift velocity of the trijunctions. In the moving frame,
\begin{eqnarray}
 \label{coupl_ani}
  \frac{\partial \delta_x}{ \partial t} &=& V \tan(\phi - \phi_0), \\ \nonumber
 	 &=&   V \tan\left(\frac{\alpha}{1 - \frac{\partial \psi}{\partial \phi} }\right),
\end{eqnarray}
since $\alpha = \delta \phi - \delta \psi = \delta \phi (1 - \frac{\partial \psi}{\partial \phi} )$.
Under the assumption that the argument inside the tangent function remains close to $0$ for all $\phi_R$
(valid in the limit of small front slopes), the expression may be linearized to yield
\begin{eqnarray}
 \label{coupl_anicou}
  \frac{\partial \delta_x}{ \partial t} &\approx&  - V \frac{\partial \zeta}{\partial x} \frac{1}{1 - \frac{\partial \psi}{\partial \phi}}, \\ \nonumber
 	 &\approx&  - V \frac{\partial \zeta}{\partial x} A_{SPA}(\phi_R).
\end{eqnarray}

For isotropic interfaces, these equations may be combined to yield an evolution equation for the local
spacing. However, for anisotropic interfaces, the inclination angle $\phi$ provides a supplementary
degree of freedom. Since both $\lambda$ and $\phi$ can be expressed in terms of $\zeta$, here
it is more convenient to write an equation for $\zeta(x,t)$ rather than $\lambda(x,t)$.

Expanding the expression of the undercooling around the homogeneous underformed state of spacing $\lambda_0$ and angle $\phi_0$, we obtain
\begin{equation}
\Delta T = \Delta T_0 + \left.\frac{\partial \Delta T}{\partial\lambda}\right|_{\lambda_0,\phi_0}(\lambda-\lambda_0)
                                   + \left.\frac{\partial \Delta T}{\partial\phi}\right|_{\lambda_0,\phi_0}(\phi-\phi_0).
\label{expanDT}
\end{equation}
The deviation of the angle is replaced by
\begin{equation}
\phi-\phi_0 = -A_{SPA}(\phi_R) \frac{\partial \zeta}{\partial x}.
\label{eq:phizeta}
\end{equation}
Taking the time derivative of Eq.~(\ref{eq:zeta}) and injecting the linear expansion Eq.~(\ref{expanDT}), one has
\begin{equation}
-G  \frac{\partial \zeta}{\partial t}=  \left.\frac{\partial \Delta T}{\partial\lambda}\right|_{\lambda_0,\phi_0}\frac{\partial \lambda}{\partial t}
                                 + \left.\frac{\partial \Delta T}{\partial\phi}\right|_{\lambda_0,\phi_0} \frac{\partial \phi}{\partial t}.
\end{equation}
Finally, replacing $\partial_t\lambda$ with Eqs.~(\ref{eq:dldt}) and (\ref{coupl_anicou}), and $\partial_t\phi$
with the time derivative of Eq.~(\ref{eq:phizeta}), one obtains the linear and homogeneous partial derivative equation (PDE) 
\begin{equation}
\frac{\partial}{\partial t} \left(\zeta - \ell_0 \frac{\partial \zeta}{\partial x} \right)= D_0 \frac{\partial^2 \zeta}{\partial x^2}
\label{eq:zetaevol}
\end{equation}
with a diffusion coefficient
\begin{eqnarray}
D_0 &=& \frac{\lambda_0 V}{G} A_{SPA}(\phi_R) \left.\frac{\partial \Delta T}{\partial\lambda}\right|_{\lambda_0,\phi_0}, \\ \nonumber
    &=& \frac{\lambda_0 V^2 K_1}{G} A_{SPA}(\phi_R) \left( 1 - \frac{1}{\Lambda_0^2(\phi_R)} \right), 
\end{eqnarray}
with $\Lambda_0(\phi_R)=\lambda_0/\lambda_m(\phi_R)$, and a length scale related to the anisotropy,
\begin{eqnarray}
\ell_0 &=& \frac{1}{G} A_{SPA}(\phi_R)  \left.\frac{\partial \Delta T}{\partial\phi}\right|_{\lambda_0,\phi_0}, \\  \nonumber
       &=&  \frac{1}{G\lambda_0} A_{SPA}(\phi_R) \left.\frac{\partial K_2}{\partial\phi}\right|_{\phi_0}.
\end{eqnarray} 

Again, it is useful to examine the specialization of these expressions to the symmetric eutectic alloy. We have
\begin{equation}
\label{D0_sym}
D_0 =  \frac{2 P(1/2) \lambda_0 V \ell_T}{\ell_D} A_{SPA}(\phi_R) \left( 1 - \frac{1}{\Lambda_0^2(\phi_R)} \right)
\end{equation}
with $\ell_T$ is the thermal length Eq.~(\ref{FBP_6}). The anisotropic length, since with $K_2 \propto \sin(\theta_\ell) \propto |\vec{\sigma}_{\alpha \beta}|$, becomes
\begin{eqnarray}
\label{l0_sym}
\ell_0 &=&   \frac{2 \ell_T d_\ell}{\gamma_\ell \lambda_0} A_{SPA}(\phi_R) \left.\frac{\partial |\vec{\sigma}_{\alpha \beta}| }{\partial\phi}\right|_{\phi_0} \nonumber \\
    &=& \frac{2\ell_T d_\ell}{\gamma_\ell \lambda_0} \frac{\gamma'_{\alpha \beta}}{\gamma_{\alpha \beta}} |\vec{\sigma}_{\alpha \beta}|
\end{eqnarray} 
since
\begin{equation}
\label{l0_der}
\left.\frac{\partial |\vec{\sigma}_{\alpha \beta}| }{\partial\phi}\right|_{\phi_0} =  \frac{\gamma'_{\alpha \beta} (\gamma_{\alpha \beta} + \gamma''_{\alpha \beta})}{|\vec{\sigma}_{\alpha \beta}|}.
\end{equation} 

\begin{figure*}[h!]
\begin{center}
\includegraphics[width=.7\textwidth]{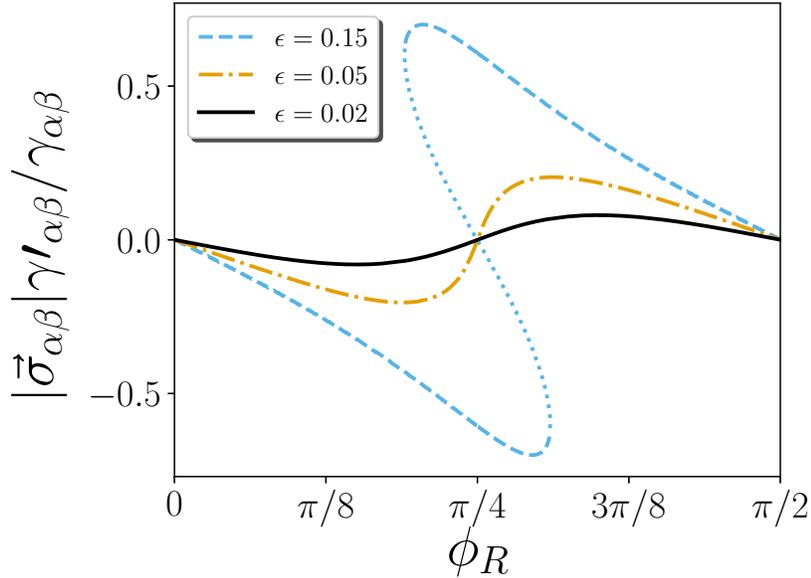}
\end{center}
\caption{Plot of $|\vec{\sigma}_{\alpha \beta}| \gamma'_{\alpha\beta}/\gamma_{\alpha\beta}$ versus $\phi_R$ for an isotropic phase diagram and under the SPA. We recall that the anisotropic length $\ell_0$ is proportional to $|\vec{\sigma}_{\alpha \beta}| \gamma'_{\alpha\beta}/\gamma_{\alpha\beta}$.
The IB energy is $\gamma_{\alpha \beta} = 1 - \epsilon \cos(4(\phi_0 - \phi_R))$ for anisotropy strengths $\epsilon=0.02$, $0.05$ and $0.15$. For $\epsilon=0.15$, the dashed curves correspond to the stable and the metastable branches, and the dotted curve corresponds to the unstable branch. }
\label{lo_lo}
\end{figure*}

We plot in Fig.~\ref{lo_lo} the function $|\vec{\sigma}_{\alpha \beta}| \gamma'_{\alpha\beta}/\gamma_{\alpha\beta}$ versus $\phi_R$ using
an IB energy $\gamma_{\alpha \beta} = 1 - \epsilon \cos(4(\phi_0 - \phi_R))$. It turns out that $\ell_0$ is a decreasing 
function of the tilting angle $\phi_0$.

The PDE for the front shape has strongly different properties for $\ell_0=0$ and $\ell_0\neq 0$.
For $\ell_0=0$ and $D_0>0$, the PDE is parabolic and reduces to the well-known diffusion equation 
which is of the same form as in the isotropic case. Indeed, the condition $\ell_0(\lambda_0,\phi_0) =0$ 
is satisfied if the temperature gradient is aligned with a direction corresponding to an extremum of 
$\gamma_{\alpha \beta}$, (\textit{i.e.} $\gamma'_{\alpha \beta}(\phi_0-\phi_R)=0$ ). This situation 
hence yields the well-known phase diffusion equation for $\lambda$, 
\begin{equation}
 \label{eq_LambD30}
 \frac{\partial \lambda}{ \partial t} =   D_0 \frac{\partial^2 \lambda}{ \partial x^2}
\end{equation}
with $D_0 = A_{SPA}(\phi_R) D_\lambda = A_{SPA}(\phi_R) \frac{\lambda_0 V}{G} \left.\frac{\partial \Delta T}{\partial\lambda}\right|_{\lambda_0,\phi_0}$. That is, the behavior of such a front is identical to the one of an
isotropic front, but with a phase diffusion coefficient that is multiplied by the anisotropy function $A_{SPA}$.
For evolution around a minimum of $\gamma_{\alpha \beta}$, the anisotropy function is lower than unity,
and the dynamics of spacing relaxation is slowed down. Conversely, when the front evolves around a maximum
of  $\gamma_{\alpha \beta}$ the dynamics is accelerated.
  
For this case, any long-wavelength and small amplitude perturbation (\textit{i.e.}, for $k\lambda_0\ll 1$ and $\delta \ll \lambda_0$) of the form
\begin{equation}
 \label{eq_pert}
 \lambda(x,t) =   \lambda_0 [1+ \delta \exp(i k x +\omega_k t)]
\end{equation}
will grow with a rate
\begin{equation}
 \label{eq_omepert}
  \omega_k = -D_0 k^2.
\end{equation}
Therefore, according to Eq.~(\ref{eq_omepert}) the system is stable if $D_0>0$, and unstable otherwise. 
One deduces that, as long as the anisotropy function $A_{SPA}$ remains positive, the threshold of stability is given by $\Lambda_0(\phi_R)=\lambda_0/\lambda_{m}(\phi_R)=1$, which is the same as for isotropic IB energy
(up to the different expression of $\lambda_m$).

In contrast, for $\ell_0\neq0$, the term $\ell_0 \partial_t\partial_x \zeta$ in the evolution equation Eq.~(\ref{eq:zetaevol}) breaks the parity symmetry $x \rightarrow -x$.
The PDE is hyperbolic, and therefore corresponds to a wave-propagation equation, see appendix \ref{APP_D} for details.
In both cases, $\ell_0=0$ and  $\ell_0\neq0$, the PDE reduces to an initial values problem.

\subsection{Normal Mode Analysis}

In order to obtain the normal mode analysis we write the Fourier representation of the  front shape $\zeta(x,t)$ as
\begin{equation}
\zeta(x,t) = \frac{1}{\sqrt{2\pi}} \int_{-\infty}^{+\infty} A_k(t) \exp(-ikx) \mathrm{d}k 
\label{eq:zetafour}
\end{equation}
with $k$ the wave number and $A_k(t)$ the amplitude of the mode $k$.
In reciprocal space, the PDE for $\zeta(x,t)$ given by Eq.~(\ref{eq:zetaevol}) reduces to 
an ordinary differential equation in time for the amplitudes
\begin{equation}
 \dot{A}_k(t) - \omega A_k(t) =0
\label{eq:zetafour2}
\end{equation}
with the dispersion relation $\omega= - \frac{D_0k^2}{1+i\ell_0k}$. This can be rewritten as
\begin{equation}
 \label{eq:disp}
 \omega= \omega_R + i \omega_I = \frac{-D_0k^2}{1+\ell_0^2k^2} + i \frac{D_0\ell_0k^3}{1+\ell_0^2k^2}.
\end{equation}
The phase velocity is given by $v_p = \omega_I/k= D_0\ell_0k^2/(1+\ell_0^2k^2)$
and the group velocity by $v_g = \mathrm{d} \omega_I /\mathrm{d}k= D_0\ell_0k^2(3+\ell_0^2k^2)/(1+\ell_0^2k^2)^2$.
Both of these velocities tend to the constant $D_0/\ell_0$ for 
$k\rightarrow \infty$ and behave like $\sim D_0\ell_0k^2$ for $k\rightarrow 0$. 
The solution of the Fourier amplitudes reads
\begin{equation}
 A_k(t) = A_k(0) \exp(\omega t)
\label{eq:zetafour3}
\end{equation}
with $A_k(t=0)= \frac{1}{\sqrt{2\pi}} \int_{-\infty}^{+\infty} \zeta(x,t=0) \exp(ikx) \mathrm{d}k$.

Therefore, the solution corresponding to an initial state $\zeta(x,t=0)= a_0 \cos(2\pi x/\lambda_p)$, 
where $\lambda_p\gg \lambda_0$ is the wavelength of the perturbation, one gets
\begin{eqnarray}
\label{eq:zetafour4}
 \zeta(x,t) &=& \frac{a_0}{2} \int_{-\infty}^{+\infty} [\delta(k-k_p)-\delta(k+k_p)] \nonumber \\ 
  &&\times \mathrm{e}^{\omega t}\mathrm{e}^{-ikx} \mathrm{d}k, \nonumber \\ 
	    &=& a_0 \mathrm{e}^{\omega_R(k_p)t} \cos(\omega_I(k_p)t-k_px)
\end{eqnarray}
with $k_p=2\pi/\lambda_p$ the wave number of the perturbation. For this, we have used the parity 
properties $\omega_R(k) = \omega_R(-k)$ and $\omega_I(k) = -\omega_I(-k)$.
Clearly, Eq.~(\ref{eq:zetafour4}) corresponds to a time-damped or -amplified wave where the sign 
of $\ell_0$ gives the direction of propagation.

\section{Discussion}
\label{section:Disc}

Up to now, we have shown that the anisotropy of the interphase boundary changes the nature of the evolution equation for the front shape. This equation is a wave equation for anisotropic interphase boundaries, in contrast to the previously known diffusion equation in the isotropic case.
In order to investigate more quantitatively the influence of the cross term in Eq.~(\ref{eq:zetaevol}), one may compare the phase velocity in the small wave number limit, $v_p(k\rightarrow 0)\sim D_0\ell_0k^2$, to the pulling velocity $V$. For the sake of simplicity, we perform the calculations for a eutectic alloy with symmetric phase diagram; no qualitative changes are expected if this restriction is relaxed. For a symmetric eutectic alloy, one has
\begin{equation}
 \frac{v_p(k\rightarrow 0)}{V} \sim \left(2P(1/2)\lambda_0\frac{\ell_T}{\ell_D}\right)^2 A_{SPA} \frac{\gamma_{\alpha \beta}'}{\gamma_{\alpha \beta}} \left(\frac{\Lambda_0^2-1}{\Lambda_0^4} \right) k^2
 \label{eq:vprop}
\end{equation}

We examine a perturbation that has a wavelength of ten lamellar spacings, that is, $k \approx 2\pi/ (10\lambda_0)$. For typical values of the other parameters, $A_{SPA}\approx 1$, $\gamma_{\alpha \beta}'/\gamma_{\alpha \beta}\sim n \epsilon \approx 0.2$, $\Lambda_0\approx 1.1$, $\ell_T/\ell_D \approx 4$, one obtains $v_p/V \approx 8\times 10^{-4}$  and $\ell_0/\lambda_0 \approx 5\times 10^{-2}$. This means that the propagation of the wave induced by the anisotropy is very slow, and will be difficult to observe on the typical time scale of directional solidification experiments. Note that according to Eq.~(\ref{eq:vprop}) the phase velocity depends on the distance of the initial spacing from the minimum-undercooling spacing through the factor that depends on $\Lambda_0$. However, since $\Lambda_0$ typically remains close to unity, our conclusion is not limited to the particular value of $\Lambda_0$ taken in the calculation.

Since the propagation of waves is very slow, we can reasonably neglect the cross term in Eq.~(\ref{eq:zetaevol}) for the description of experiments. Therefore, the phase diffusion equation given by Eq.~(\ref{eq_LambD30}) remains a good approximation, even if the
temperature gradient is not aligned with a direction corresponding to an extremum of $\gamma_{\alpha,\beta}$ (\textit{i.e.}, for tilted interphases).

Two further comments can be added at this point.
First, the lateral propagation of patterns and the evolution of spacings have been recently studied in experiments and phase-field simulations of cellular and dendritic arrays in dilute binary alloys \cite{Song_Karma_PRM_2018},
 and an evolution equation for the spacing has been extracted, which also contains propagative and diffusive terms. However, in contrast to our findings, for dendrites the propagative term dominates over the diffusive one.
  This difference points to the very different roles that crystalline anisotropy plays for the selection of dendritic and eutectic patterns.
   Second, we have relied here on the symmetric pattern approximation. Evolution equations for the spacing have been derived directly from the free-boundary problem in the limit of high temperature gradients for eutectics \cite{Caroli_Caroli_JPhysFrance_1990},
    and by a perturbation analysis of the boundary integral equation for cells \cite{Brattkus_Misbah_PRL_1990}. While it might be possible to use similar methods for a more rigorous derivation of the front evolution equation obtained here, this would certainly be a difficult undertaking.
     Moreover, a more rigorous treatment would probably not decisively alter the order-of-magnitude estimates obtained above.

\section{Conclusion}
\label{section:Concl}

We have developed an evolution equation for the envelope of lamellar eutectic solidification fronts in two dimensions, which corresponds to experiments in thin samples, taking into account the anisotropy of the solid-solid interphase boundaries. This generalizes previous works on fronts with isotropic interfaces \cite{Jackson_Hunt_TMSA_1966,Langer_PRL_1980,Datye_Langer_PRB_1981}. By replacing Cahn's hypothesis (lamallae always grow normal to the envelope) used for isotropic systems by the Symmetric
Pattern Approximation (SPA) that is derived from the balance of torques at the trijunction points, we have demonstrated that the evolution equation contains a propagative term which involves a new characteristic length scale $\ell_0$. This is striking, because a local effect 
changes the nature of the equations describing the evolution of the system at large scales.
Moreover, the diffusive term that is already present for isotropic interfaces gets multiplied 
by a factor that depends on the anisotropy of the interphase boundaries.

A quantitative analysis of the new equation reveals that for typical directional solidification conditions, the phase velocity of the propagative modes is too slow to be observable. Therefore, 
the propagative evolution equation can be reasonably replaced by a spacing diffusion equation 
as in the isotropic case, where the spacing diffusion coefficient is multiplied by the anisotropy function $A_{SPA}$. In addition, the dependence of the minimum undercooling spacing on the tilting 
angle must also be taken into account in order to correctly evaluate the reduced initial spacing.

It should be recalled that direct experimental measurements and phase field simulations for  
lamellar eutectics with isotropic interphases \cite{Plapp_Karma_PRE_2002,Akamatsu_Karma_MMTA_2004} 
have shown that Cahn's hypothesis is not strictly valid for isotropic systems: the trijunctions 
also slightly move in the direction parallel to the envelope of the composite front in 
addition to the normal-growth conjecture, with a velocity that is proportional to the local 
gradient of the spacing. Despite the fact that this effect is small, it introduces
a stabilizing term in the diffusion equation that leads to an overstability with respect
to the theory. We expect a similar contribution in the anisotropic case, but since no
analytical description of this phenomenon is available, only numerical simulations could
permit to clarify this issue.
We hope to report on the results of such simulations in the near future.

\section{Ackowlegdements}

The authors thank S. Akamatsu, G. Faivre, and S. Bottin-Rousseau
for many useful discussions. This research was supported by the
Centre National d’Études Spatiales (France) and by the ANR 
ANPHASES project (M-era.Net:ANR-14-MERA-0004).

\appendix

\section{Limit of Stability of $\phi_0(\phi_R)$ within the SPA}
\label{APP_B}

As mentioned in the section \ref{SPA_P}, for an underformed steady-state with a planar front, the SPA (\textit{i.e.} $\vec{\sigma}_{\alpha \beta} \cdot \hat{t}_f = 0$) imposes for 
\begin{equation}
 \label{eq_tilt0A}
  \phi_0 = - \arctan\left(\frac{\gamma_{\alpha \beta}'(\phi_0 -\phi_R)}{\gamma_{\alpha \beta}(\phi_0-\phi_R)}\right).
\end{equation}
Using the formalism of the dynamical system \cite{Manneville_1990}, the problem can be tackled by writing 
\begin{equation}
 \label{eq_tilt0_dynA}
  \frac{\partial \phi}{\partial t} = F(\phi;\phi_R)
\end{equation}
with 
\begin{equation}
 \label{eq_tilt0_dyn2A}
    F(\phi;\phi_R)= \phi +\arctan\left(\frac{\gamma_{\alpha \beta}'(\phi -\phi_R)}{\gamma_{\alpha \beta}(\phi-\phi_R)}\right).
\end{equation}
a nonlinear function of $\phi$. The dynamics is fully determined by the nature and the position of the fixed points of $F$ given by
\begin{equation}
 \label{eq_tilt0_fixA}
 F(\phi_0;\phi_R) = 0.
\end{equation}
The problem reduces to solve how the fixed points $\phi_0$ depends on $\phi_R$ (seen as a control parameter).
Therefore, the fixed points have to be viewed like implicit functions of $\phi_R$.
Interestingly, for a $n-$fold function for $\gamma_{\alpha \beta}$, the function $F(\phi_0;\phi_R)$ is symmetric with respect to the transformation $\phi_R \rightarrow 2\pi/n- \phi_R $ and $\phi_0 \rightarrow -\phi_0$.

Furthermore, applying the transformation around a fixed point $\phi_R \rightarrow \phi_R +\delta \phi_R$
and $\phi_0 \rightarrow \phi_0 +\delta \phi_0$,  Eq.~(\ref{eq_tilt0A}) leads to the relation
\begin{eqnarray}
 \label{eq_undef1A}
 \delta \phi_0 &=& \frac{\partial F/\partial \phi_R}{\partial F/\partial \phi_0} \delta \phi_R,\\ \nonumber
               &=& \frac{\gamma''_{\alpha \beta} \gamma_{\alpha \beta} - \gamma^{'2}_{\alpha \beta}}{\gamma_{\alpha \beta}(\gamma_{\alpha \beta} +\gamma''_{\alpha \beta})} \delta \phi_R.
\end{eqnarray}
The limit of stability corresponds to the turning points given by the
conditions $\frac{\partial F}{\partial \phi_R} \neq 0$ and $\frac{\partial F}{\partial \phi_0} = 0$
or equivalently $\gamma_{\alpha \beta} +\gamma''_{\alpha \beta}=0$.

Unfortunately, this local investigation of the stability does not allow to obtain the limit between the stable 
and the meta stable branches. However, using symmetric consideration, those points are given by
$\{\phi_R=2\pi/n,\phi_0(2\pi/n)\}$.

\section{Transformation of the Equation of Evolution}
\label{APP_D}
As written in the core of the text, 
the Partial Derivative Equation (PDE)
\begin{equation}
\frac{\partial}{\partial t} \left(\zeta - \ell_0 \frac{\partial \zeta}{\partial x} \right)= D_0 \frac{\partial^2 \zeta}{\partial x^2}
\label{eq:zetaevol2}
\end{equation}
governing the evolution of the average front has an unusual form from a physicist point of view. In this appendix, we want to demonstrate that, doing a suitable change of variables, one can write this PDE under a more common form without cross derivative and allowing to extract a characteristic time.  
The determinant of the characteristic polynomial of Eq.~(\ref{eq:zetaevol2}) is $\Delta_\zeta = \ell_0^2$ which leads to the roots $r_1=0$ and $r_2=\ell_0/D_0$.

One introduces the characteristic coordinates $\xi = t - r_1 x = t$ and $\eta = t - r_2 x = t -\ell_0 x/D_0$.
Using those variables, the linear differential operator becomes symmetric and  Eq.~(\ref{eq:zetaevol2}) reads
\begin{equation}
\frac{\ell_0^2}{D_0} \partial_{\xi \eta}\zeta + \partial_{\xi}  \zeta + \partial_{\eta}  \zeta = 0. 
\label{eq:zetacano}
\end{equation}
Finally, in order to remove the cross derivative, one sets $\alpha = \xi +\eta=2t-\ell_0x/D_0$ and $\beta=\xi-\eta=\ell_0x/D_0$, one obtains
\begin{equation}
\partial_{\alpha \alpha}\zeta - \partial_{\beta \beta}  \zeta + \frac{1}{\tau_{ab}} \partial_{\alpha}  \zeta = 0.
\label{eq:zetacano2}
\end{equation}
with $\tau_{ab} = \frac{\ell_0^2}{2D_0} $ a characteristic time.

\end{document}